\documentclass{nsart_eng_new}
\usepackage{cite}
\usepackage{graphicx}
\usepackage{amsmath,amssymb}
\usepackage{siunitx}

\DeclareSIUnit\angstrom{\text{\AA}}
\newcommand{\Dcoll}{\Delta E_{\mathrm{coll}}}
\newcommand{\Dnucl}{\Delta E_{\mathrm{nucl}}}
\newcommand{\EAFM}{E_{\mathrm{AFM}}}
\newcommand{\Esp}{E_{\mathrm{sp}}}

\begin{document}

\title{Skyrmions in Synthetic Antiferromagnets: Collapse and Nucleation}

\author[M.~N.~Potkina]{M.~N.~Potkina}

\address{Infochemistry Scientific Center, ITMO University, 197101 St. Petersburg, Russia}


\begin{abstract}
Magnetic skyrmions in synthetic antiferromagnets are promising nanoscale bits, but their
usefulness depends on how reliably a written pair survives and can be created.
Using a reduced lattice model, we compute minimum energy paths for collapse of an antiferromagnetically bound
skyrmion pair and for reverse nucleation from a pinned antiferromagnetic reference state.
With antiferromagnetically pinned boundaries, the main saddle energy changes only weakly with pinned-island size,
whereas the skyrmion-pair minimum carries a strong size-dependent boundary penalty.
For large pinned islands, collapse is layer-sequential and can pass through a single-layer skyrmion
intermediate whenever this state satisfies the relaxation criterion.
The much larger reverse barrier for nucleation shows a strong asymmetry with collapse in the same
pinned-boundary model and is consistent with assisted layer-sequential writing.
\end{abstract}

\maketitle

\section{Introduction}

Magnetic skyrmions are localized chiral spin textures that can be moved by spin-polarized currents and are
therefore considered as information carriers for racetrack-type memory and logic devices
\cite{Wiesendanger16,Finocchio16,Fert17}.
For memory applications, however, miniaturization creates a direct tension between bit density and thermal
stability.
Nanometer-scale skyrmions in ultrathin ferromagnetic films can have low activation barriers against collapse,
whereas larger skyrmions stabilized by dipole--dipole interactions are less attractive for dense memory because
stray fields increase their size and mediate unwanted long-range interactions between neighboring bits
\cite{Buttner18}.

Antiferromagnetic and compensated ferrimagnetic systems offer a natural way to reduce these drawbacks.
In compensated materials the net magnetization and dipolar fields are small, and the cancellation of the
gyrotropic response suppresses the skyrmion Hall effect.
Fast current-driven motion of domain walls and small skyrmions has been demonstrated in compensated
ferrimagnets \cite{Caretta18}.
Transition-state-theory calculations for ideal antiferromagnetic films have shown that, in the absence of
magnetostatic effects, the activation barriers and entropic factors for skyrmion collapse are closely related
to those in corresponding ferromagnetic films, while the dynamics can modify the prefactor of the rate
\cite{Potkina20}.

A practical difficulty in intrinsic antiferromagnets is detecting and controllably manipulating individual
topological textures.
This has motivated the use of synthetic antiferromagnets (SAFs), where two ferromagnetic layers are separated
by a nonmagnetic spacer that mediates antiferromagnetic Ruderman--Kittel--Kasuya--Yosida (RKKY) coupling
\cite{Duine20}.
The ferromagnetic layers can be optimized separately, the interlayer coupling can be tuned by the spacer, and
the two components of a coupled skyrmion pair remain accessible to layer-sensitive probes.
Early theoretical work proposed multilayer SAF racetracks as thermally robust skyrmion conduits with
suppressed transverse motion \cite{Zhang16}.
Bilayer theory further showed that interlayer coupling can bind skyrmions in the two films and modify their
collective response compared with isolated ferromagnetic skyrmions \cite{Koshibae17}.
Artificial SAF multilayers also provide a tunable platform in which the effective coupling and field balance
can be adjusted by the multilayer design \cite{Li20}.
Subsequent experiments demonstrated room-temperature formation and current-induced motion of synthetic
antiferromagnetic skyrmion bubbles with negligible skyrmion Hall deflection \cite{Dohi19}, room-temperature
stabilization of antiferromagnetically coupled skyrmions of about ten nanometers in Pt/Co/Ru-based SAFs
\cite{Legrand20}, and zero-field skyrmions in uncompensated SAF multilayers \cite{Chen_20}.
Material-specific modeling of Co/Pt-based multilayers has clarified how DMI, anisotropy, layer thickness, and
interlayer exchange select FM- or AFM-coupled skyrmion states \cite{Jia_20}.
More recent experiments have shown local nucleation by current injection and ultrafast laser pulses
\cite{Juge_22} and current-induced motion in compensated SAFs with velocities up to hundreds of meters per
second \cite{Pham24}.
These results establish SAF skyrmions as a realistic platform for memory and logic devices.

The writing problem is more challenging than controlling motion.
Several assisted writing protocols have already been proposed or demonstrated.
Local current injection and ultrafast laser excitation can nucleate SAF skyrmions \cite{Juge_22}.
Voltage-controlled magnetic anisotropy has been proposed for writing a skyrmion at a chosen position in an SAF
racetrack \cite{Qiu22}.
Thermal assistance has been used to control skyrmion generation in SAFs \cite{Chen_22}.
In single-layer skyrmion tracks, nonmagnetic impurities and engineered defects can also affect skyrmion memory
operation \cite{PotLobUzd20}.
MEP calculations have shown, in particular, that notches and linear vacancy defects can lower nucleation
barriers \cite{PotLob24,PotLob25}.
Hybrid ferromagnet/SAF geometries have also been proposed, where a skyrmion created in a ferromagnetic section
enters the SAF section and generates a corresponding skyrmion in the second layer through interlayer exchange
coupling \cite{Deng23}.
Together, these works make assisted or sequential writing technologically plausible.
The question here is whether a layer-by-layer route is only a convenient dynamical protocol, or also a feature
of the energy landscape of a pinned SAF island.

The thermal stability of skyrmions is determined by rare transitions between local minima of the magnetic
energy.
Minimum energy path (MEP) calculations and harmonic transition state theory (HTST) provide a microscopic way
to find the relevant saddle points and activation barriers \cite{Lobanov21_PJETP,Lobanov21_CPC}.
A complete thermal-stability estimate would additionally require the dynamical prefactor, Hessian
determinants, and conversion of the reduced energy scale to the physical multilayer.
For single-layer skyrmions, related MEP studies have shown that size, texture shape, and field-dependent shifts
of metastable and saddle-point energies all enter the activation barrier
\cite{Varentsova18,PotLobUzd23}.
For SAFs, previous calculations have shown that interlayer exchange can change both the collapse barrier and
the collapse mechanism \cite{Schrautzer22,Voronin22}.
These studies established that interlayer coupling can produce non-simultaneous, layer-resolved collapse.
The new elements addressed here are the behavior of this mechanism in pinned islands, the
lateral-size dependence of the interior-collapse barrier separated into saddle and pair-minimum contributions,
the connection between collapse and the reverse
layer-sequential nucleation path on the same MEP, and the stability map of the independently relaxed
upper-layer single-skyrmion intermediate.
However, for a skyrmion used as a memory bit, two distinct questions must be separated.
The first is barrier-limited retention: what collapse barrier protects an already written skyrmion pair
against thermally activated decay?
The second is writability: if the pinned collinear antiferromagnetic state is the unwritten reference state,
what route can create the skyrmion pair without relying on exponentially rare spontaneous nucleation?

Here we calculate MEPs for collapse and nucleation of bound skyrmion pairs in pinned SAF islands using
a reduced lattice model parameterized, for the retained interaction terms, from the length and energy scales of
the Legrand-type SAF stack \cite{Legrand20}.
The experimental stack used for this parameterization consists of two magnetic layers separated by a
nonmagnetic metal layer.
The effective exchange interaction between magnetic layers, described by the Ruderman--Kittel--Kasuya--Yosida
(RKKY) mechanism, has a strength and sign that depend on the thickness of the nonmagnetic spacer.
This thickness is chosen so that the interlayer interaction is antiferromagnetic.
The magnetic layers are Co films interfaced with the heavy metal Pt, which induces Dzyaloshinskii--Moriya
antisymmetric exchange in the magnetic subsystem.
To obtain spatially localized bound skyrmion pairs in the SAF, additional
[Pt(0.45 nm)/Co(0.6 nm)]$_4$ magnetic layers with strong perpendicular uniaxial anisotropy were introduced
into the system.
These layers induce a fixed effective bias field $\mathbf B_b$ via the Pt spacer in the nearest magnetic layer
of the SAF.

The model retains intralayer exchange, interfacial DMI, antiferromagnetic RKKY coupling, and Zeeman terms for
the applied field and the fixed effective bias field, which are the interactions most directly involved in the
collapse path studied below.
It is a reduced energy-landscape model rather than a full micromagnetic description of the experimental
multilayer: perpendicular anisotropy, magnetostatic/dipolar fields, and dynamical prefactors are not included
explicitly.
Antiferromagnetic compensation in SAFs is expected to reduce the net moment and the long-range stray field,
which is one reason such structures are attractive for dense skyrmion devices \cite{Duine20,Zhang16}.
Residual demagnetizing fields from unequal layer textures, finite layer separation, and edges can still matter,
especially in asymmetric multilayers \cite{Correia24}.
The calculations therefore identify mechanisms and trends within the retained interactions; quantitative
skyrmion radii, stability windows, activation barriers, and lifetimes for a specific Legrand- or Juge-type
stack require the omitted terms and prefactors.

Here the pinned island has a limited role: it lets us estimate the saddle point for an interior collapse event
in an infinite or sufficiently large collinear SAF region, provided the saddle is localized away from the
artificial boundary.
For such a localized saddle, increasing the island size gives a direct convergence check, and the saddle
energy is the quantity the method is designed to estimate accurately.
The boundary correction to the energy of the skyrmion-pair minimum is treated only approximately.
Since fixed boundary spins impose an additional constraint on the relaxed texture, the calculated pair energy
should be regarded as an upper-bound estimate for the corresponding skyrmion-pair branch when boundary
relaxation is allowed.
The reported barriers combine a localized interior-saddle estimate with a boundary-dependent
estimate of the skyrmion-pair minimum.
This distinction explains the size trend: changes in the pinned-boundary barrier can originate from
the boundary penalty to the skyrmion-pair energy rather than from a comparable shift of the interior saddle.
Accordingly, ``barrier'' below means the calculated pinned-boundary MEP-channel barrier unless explicitly
stated otherwise.

Within this reduced model, the MEPs lead to three main conclusions.
First, for the antiferromagnetically pinned boundary used here, the calculated interior saddle energy varies
only weakly with pinned-island size, while the energy of the skyrmion-pair minimum changes strongly.
For the parameters studied below, the pinned-boundary collapse barrier of this channel is almost zero in the
smallest pinned island, reaches several exchange energies for intermediate sizes, and becomes largest for the largest islands
considered because it is measured as $\Dcoll=\Esp-E_{\mathrm{pair}}$.
The applied field produces only a secondary change over the explored range.
Second, the collapse is generally not simultaneous in the two SAF layers.
For the field orientation used here, the lower-layer skyrmion collapses first and the path can pass through a
single-skyrmion state in the upper layer.
This upper-layer state satisfies the relaxation-based local-minimum criterion only in part of the size-field
diagram; when it is absent, the collapse becomes effectively single-stage.
Third, the reverse nucleation barrier from the pinned collinear antiferromagnetic reference state is much
larger than the collapse barrier of an already written pair.
Within the same reduced barrier picture, this makes spontaneous formation along the calculated path unlikely but
also makes direct thermal nucleation an inefficient writing route.
This points to layer-sequential writing: if the upper-layer skyrmion is created or injected first, interlayer
exchange can complete the bound SAF pair through a lower second-step barrier.

\section{Model and method}

We use a lattice spin model representing two ferromagnetic layers of an SAF separated by a nonmagnetic spacer.
Each lattice site carries a unit vector $\mathbf S_i^{(l)}$, where $l=1,2$ labels the two magnetic layers.
The intralayer interactions include ferromagnetic exchange and Dzyaloshinskii--Moriya interaction (DMI).
The two layers are coupled antiferromagnetically through an RKKY-type interlayer exchange.
Following the experimental geometry of Ref. \cite{Legrand20}, a third, uniformly magnetized bias layer is not
treated dynamically; its effect is included as a fixed effective bias field acting on the lower SAF layer.

The reduced Hamiltonian is written as \cite{Voronin22}

\begin{align}
\label{eq:Hamiltonian}
\mathcal{E} = &- \sum_{l=1}^2 \sum_{\langle i,j\rangle}
\left[ J\,\mathbf{S}_i^{(l)}\cdot \mathbf{S}_j^{(l)}
+ \mathbf{D}_{ij}\cdot \left(\mathbf{S}_i^{(l)}\times \mathbf{S}_j^{(l)}\right)\right]
\nonumber \\
&-J_{12}\sum_i \mathbf{S}_i^{(1)}\cdot \mathbf{S}_i^{(2)}
-\mu \sum_i \left[(\mathbf B_b+\mathbf B)\cdot\mathbf S_i^{(1)}
+\mathbf B\cdot\mathbf S_i^{(2)}\right].
\end{align}

Here $J>0$ is the intralayer exchange constant.
For a nearest-neighbor bond directed from site $i$ to site $j$,
$\mathbf e_{ij}=(\mathbf r_j-\mathbf r_i)/a$, the interfacial DMI vector is
\begin{equation}
\mathbf D_{ij}=D\left(\mathbf e_{ij}\times\mathbf e_z\right),
\qquad D>0.
\end{equation}
With this convention, bonds along $+\mathbf e_x$ and $+\mathbf e_y$ have
$\mathbf D_{ij}=(0,-D,0)$ and $(D,0,0)$, respectively, as in the numerical model.
The interlayer coupling is antiferromagnetic, $J_{12}<0$.
The vector $\mathbf B=B\mathbf e_z$ is the applied field, identical for both SAF layers.
The vector $\mathbf B_b=B_b\mathbf e_z$ is the fixed effective bias field from the additional layer, acting
only on the lower layer ($l=1$).
Thus the net field entering the Zeeman term is $\mathbf B_b+\mathbf B$ in layer 1 and $\mathbf B$ in layer 2.
Equation \eqref{eq:Hamiltonian} is a reduced Hamiltonian.
It keeps the local exchange, interfacial DMI, interlayer RKKY coupling, applied-field Zeeman term, and fixed
effective bias field that enter directly into the collapse pathway considered below.
The omitted terms include a separate perpendicular anisotropy, $-K\sum_i(S_{i,z}^{(l)})^2$, and the long-range
dipole-dipole interaction.
The stabilizing influence of anisotropy and magnetostatics is therefore not resolved explicitly; instead, the
calculation assumes that the skyrmion-pair and single-layer-skyrmion states can be relaxed for the chosen
effective parameters.
Consequently, the absolute stability windows, skyrmion radii, collapse paths, activation barriers, and
lifetimes are model-level quantities rather than one-to-one predictions for the full multilayer.

The mapping from discrete parameters to micromagnetic ones is controlled by dimensionless ratios.
For a square discretization with lattice constant $a$ and magnetic layer thickness $d$, the intralayer
exchange scale can be written as
\begin{equation}
J = 2d\mathcal A,
\end{equation}
where $\mathcal A$ is the micromagnetic exchange stiffness.
The DMI ratio is
\begin{equation}
\frac{|D_{ij}|}{J}=\frac{a\mathcal D}{2\mathcal A},
\end{equation}
where $\mathcal D$ is the interfacial DMI density.
The interlayer RKKY coupling gives
\begin{equation}
\frac{|J_{12}|}{J}=
\frac{a^2\mathcal J_{\mathrm{RKKY}}}{2d\mathcal A},
\qquad J_{12}<0,
\end{equation}
where $\mathcal J_{\mathrm{RKKY}}$ is the interfacial RKKY energy density.
For the fixed-bias Zeeman term, the corresponding dimensionless parameter is
\begin{equation}
\frac{\mu B_b}{J}=\frac{a^2 M_s B_b}{2\mathcal A},
\end{equation}
where $M_s$ is the saturation magnetization.
Because the RKKY coupling is interfacial, its ratio to $J$ scales as $1/d$.
This distinction is important for device interpretation: changing the magnetic thickness can change both the
absolute energy scale and the effective coupling ratios, whereas changing the lateral size $S$ at fixed
parameters mainly probes finite-size confinement.
In this sense, parameterization from the Legrand-type SAF stack means that the length scale and the coefficients
of the retained interaction terms are chosen from the corresponding micromagnetic and experimental scales.
The complete Legrand multilayer also contains perpendicular anisotropy, magnetostatic fields,
thickness-dependent material parameters, and the detailed bias-layer structure, all of which contribute to
skyrmion stabilization.

The parameters of the retained terms were taken from the continuous-model scales and experimental data of Ref.
\cite{Legrand20}: 
$a=0.2505 \, \text{nm}, \,d=1.47 \, \text{nm}$,
$\mathcal{A} = 10 \cdot 10^{-12}$ J/m,
$\mathcal{D} = 0.5 \cdot 10^{-3}$ $J/m^2$,
$\mathcal{J}_{\mathrm{RKKY}} = 0.23 \cdot 10^{-3}\,\mathrm{J\,m^{-2}}$,
$M_s = 1.2 \cdot 10^6$ A/m,
$B_b=50 \,\text{mT}$,
which gives the following set of discrete-model parameters:
$J = 183.5 \,\text{meV},\, D/J = 6.3 \cdot 10^{-3},\, |J_{12}|/J = 5 \cdot 10^{-4},\,
\,\mu = 12\mu_B$.
The pinned islands are represented as square regions on square lattices, with linear size $S$ in lattice units.
The MEP barrier calculations use $S=100$, $300$, and $500$, with applied fields
$B=-25,0,10,20,30$ mT.
With $a=0.2505$ nm, these three sizes correspond to lateral lengths of approximately $25$, $75$, and $125$ nm,
respectively.
The stability diagram of the upper-layer intermediate in Fig. \ref{fig:barrier_n}(b) is obtained from an
additional grid of independent relaxations with $S=100,200,300,400,500$ and
$B=-20,-10,0,10,20,30$ mT.
The boundary spins are pinned in the antiferromagnetic state and kept fixed throughout the calculation.
The pinned sites form the outermost lattice contour of the island; all interior spins are free to
relax.
Such a boundary condition is best viewed as a pinned island embedded in an extended collinear
antiferromagnetic background, or as an island coupled to a rigid antiferromagnetic surrounding.
It fixes the far-field background and focuses the calculated MEP on interior collapse; open-edge deformation,
skyrmion escape, and edge-assisted annihilation paths are outside this boundary-value problem.
The fixed edge also suppresses DMI-induced twisting of the homogeneous reference state.
Thus $\EAFM$ below is the energy of the pinned collinear reference state.
Pinning changes the absolute energies of finite-size skyrmion states because the texture cannot relax freely at
the boundary.
This effect is expected to be strongest for the smallest pinned island, where the skyrmion-pair texture
occupies a large fraction of the island.
Since pinning restricts the variational space, the relaxed skyrmion-pair energy obtained in this way is an
upper-bound estimate for the corresponding pair state when boundary relaxation is allowed.
To the extent that the main saddle is more localized in the interior than the relaxed pair, its energy relative
to the pinned collinear background is less sensitive to the missing edge relaxation.
This is the intended interpretation of $\Esp$: an estimate of the interior-collapse saddle in a fixed extended
antiferromagnetic background, not a free-edge saddle of an isolated island.
Because $E_{\mathrm{pair}}$ is raised by the boundary constraint, $\Dcoll=\Esp-E_{\mathrm{pair}}$ tends to
underestimate the interior-channel barrier.
Competing open-edge routes may have smaller barriers and are separate problems.
The size dependence reported below is therefore the lateral-confinement trend of this pinned-boundary
interior-collapse channel.
For each MEP parameter set we separately relax the homogeneous antiferromagnetic state, the bound skyrmion
pair, and an initial state with a skyrmion only in the upper layer ($l=2$).
These single-configuration relaxations are performed by nonlinear conjugate gradient minimization on the
fixed-length spin manifold until the maximum tangent-space energy gradient is below $10^{-9}$ in units of the
exchange energy $J$.
Under this relaxation, the homogeneous antiferromagnetic state remains unchanged within numerical precision and
is used as the pinned collinear reference state in the MEP analysis.
For the field orientation used here, the optimized collapse paths first lose the lower-layer skyrmion
($l=1$), so the stationary intermediate, when it exists, is the upper-layer skyrmion.
The lower-layer-only configuration is not a separate intermediate on the reported MEPs.
The independently relaxed upper-layer state is used to test the local stability of intermediate images by this
relaxation criterion.
MEPs connecting the skyrmion pair and the homogeneous state are obtained with a constrained MEP algorithm for
fixed-length magnetic moments \cite{Bessarab15,Lobanov_17,Lobanov21_CPC}.
The path is represented by a sequence of spin configurations, or images, on the product of unit spheres.
The end images are the relaxed local minima; they are kept fixed.
The initial path is generated by linear interpolation in the embedding spin-vector space, followed by
renormalization of each spin to unit length.
The number of images is increased with the accumulated Euclidean length of the path, which grows with pinned-island
size.
The final optimized chains used in the figures contain $20$, $23$--$37$, and $27$--$37$ images for
$S=100$, $300$, and $500$, respectively, including the two fixed endpoint images.

At each iteration the energy gradient of every free spin is projected onto the local tangent plane of the unit
sphere, and the path tangent is estimated from neighboring images using the energy-weighted prescription of the
climbing-image NEB method.
Moving images are relaxed using the component of this tangent-space gradient perpendicular to the path.
Images at energy maxima are treated as climbing images; for these images the tangential component is reversed
so that the image converges to the saddle point on the MEP.
Images at energy minima along the path are treated as descending images.
After every relaxation step all spins are projected back to unit length, and the moving images are redistributed
by interpolation at equal intervals of accumulated Euclidean distance.
The extrema images are kept fixed during this reparametrization, as in the constrained implementation of
Ref. \cite{Lobanov_17}.

The distance between two images $p_m$ and $p_n$ is the Euclidean distance in the embedding spin-vector space,
\begin{equation}
\ell(p_m,p_n)=
\left[
\sum_{l=1}^{2}\sum_{i\in\Omega}
\left|\mathbf S_{i,m}^{(l)}-\mathbf S_{i,n}^{(l)}\right|^2
\right]^{1/2},
\end{equation}
where $\Omega$ denotes the spins included in the path; fixed boundary spins make no contribution because they
are identical in all images.
The horizontal coordinate in the MEP plots is the accumulated distance
$s_n=\sum_{m<n}\ell(p_m,p_{m+1})$.

Convergence of the path is monitored by the maximum tangent-space force perpendicular to the path for moving
images and by the full tangent-space force at the climbing and descending images.
For the stationary configurations used as endpoints, saddle points, and intermediate minima, the maximum
tangent-space energy gradient after the final relaxation is below $10^{-9}$ in units of the exchange energy
$J$.
The number of images and relaxation iterations were increased until the saddle energies and the presence or
absence of the single-layer minimum no longer changed at the energy resolution used for the path analysis.
All reported barriers are evaluated from the optimized discrete images, not from the interpolating curves used
only for plotting.
Candidate saddle points are the final climbing images at strict local maxima of the discrete energy profile.
Their position is checked by an additional reparametrization of the neighboring moving images; the images on
the two sides must relax toward the adjacent minima or endpoints.
The upper-layer single-skyrmion state is counted as locally stable only when the corresponding path image is a
strict discrete minimum, relaxes independently by nonlinear conjugate gradient to a stationary configuration
with maximum tangent-space gradient below $10^{-9}J$, and is separated from adjacent saddle configurations on
the optimized path.
For the accepted intermediate minima in the final data, the smallest energy decrease relative to the neighboring
images is $8\times10^{-4}J$; shallower features below $5\times10^{-4}J$ are treated as numerical path
resolution rather than as separate metastable states.
No Hessian spectra were evaluated in the present work.
Consequently, the stability classification is based on nonlinear relaxation and discrete MEP structure rather
than on a positive-definite Hessian test.
The saddle terminology refers to the climbing-image stationary points on the optimized MEP, while quantitative
HTST prefactors are outside the scope of the calculation.

Two activation barriers are useful.
The collapse barrier of an already written skyrmion pair is
\begin{equation}
\Dcoll = \max_{\mathrm{path}} E - E_{\mathrm{pair}},
\end{equation}
where $E_{\mathrm{pair}}$ is the energy of the bound pair.
The reverse barrier from the pinned collinear antiferromagnetic reference state is
\begin{equation}
\Dnucl = \max_{\mathrm{path}} E - \EAFM.
\end{equation}
Throughout this paper, the terms ``collapse'' and ``nucleation'' refer to thermally activated transitions on
the static energy landscape of Eq. \eqref{eq:Hamiltonian}, with no nonconservative driving force.
Thus $\Dnucl$ is an equilibrium reverse MEP barrier for spontaneous formation along this path.
Current-, voltage-, laser-, or geometry-assisted writing is a nonequilibrium process and need not follow the
reverse static MEP; the reverse path calculated here provides a reference energy scale and identifies possible
intermediate states for such driven protocols.
In the pinned-boundary MEPs below, the relaxed homogeneous state is lower in energy than the calculated pair
minima, so $E_{\mathrm{pair}}-\EAFM>0$ and $\Dnucl$ is much larger than $\Dcoll$.
These statements are conditioned on the pinned-boundary states and reference energy defined above.
The optimized MEPs reported below define one interior collapse/nucleation channel for this boundary-value
problem; random, spiral, edge-twisted, and edge-escape routes were not searched for competing lower barriers.

\section{Results}

\subsection{Layer-by-layer collapse mechanism}

\begin{figure}
\centering
\includegraphics[width=\linewidth]{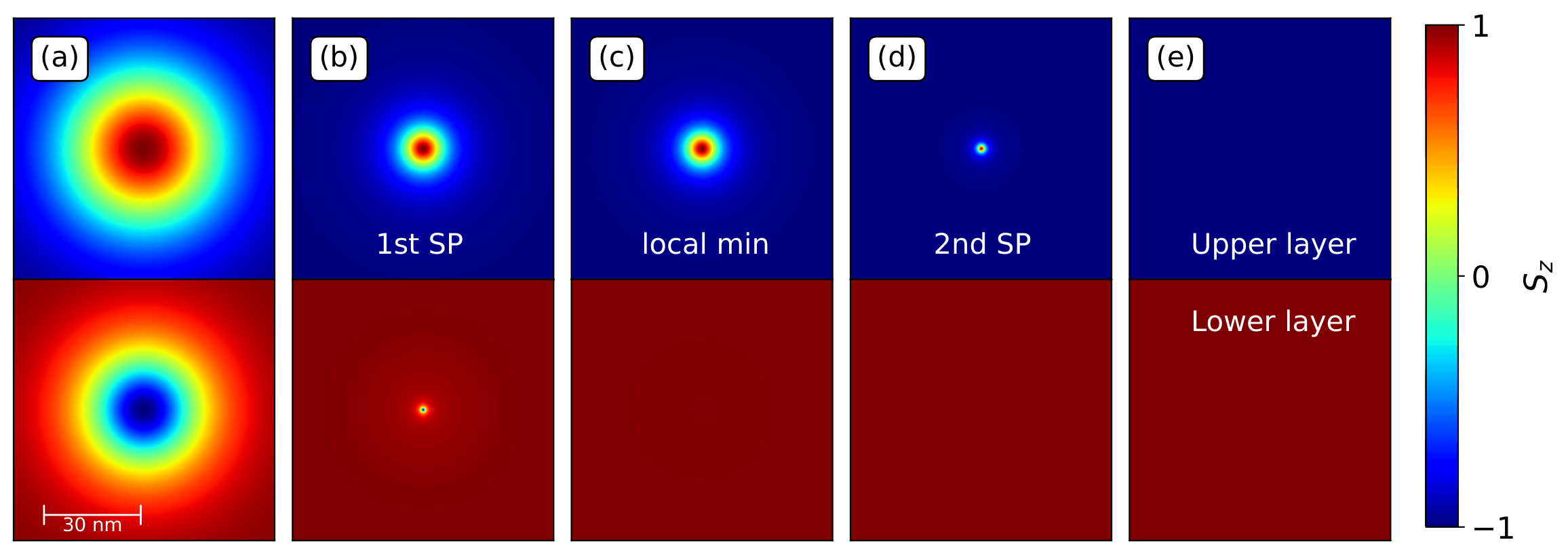}
\caption{Configurations along the minimum energy path for skyrmion-pair collapse at $B=30$ mT and $S=500$.
The color scale shows the spin component $S_z$.
The top row shows the upper layer and the bottom row shows the lower layer.
Only the central $300\times 300$ spins of the full system are shown.
The path starts from a bound skyrmion pair (a).
At the first saddle point the lower-layer skyrmion collapses (b), leaving a local minimum with a single
skyrmion in the upper layer (c).
The remaining skyrmion collapses through the second saddle point (d), and the path ends in the pinned
homogeneous antiferromagnetic reference state (e).}
\label{fig:configs}
\end{figure}

Figure \ref{fig:configs} shows representative configurations along the MEP for $B=30$ mT and $S=500$.
The initial state is a bound skyrmion pair with opposite core magnetizations in the two layers.
Skyrmions in the upper and lower layers have different sizes because the lower layer is subject to a net
field of magnitude $B_b + B = 80$ mT, opposite to the direction of magnetization at the skyrmion core, while
the upper layer is subject only to the applied field of magnitude $B = 30$ mT, parallel to the magnetization at
the center.
Increasing the applied field decreases the skyrmion size in the lower layer but increases it in the upper
layer.
When the applied field is set to $B = -B_b/2$, the skyrmion radius in both layers is the same.
The collapse is not simultaneous.
First, the skyrmion in the lower layer collapses through the higher saddle point, while the skyrmion in the
upper layer remains localized.
The path then relaxes to a local minimum corresponding to a single skyrmion in the upper layer.
The second, lower saddle point corresponds to the collapse of this remaining skyrmion, after which the system
reaches the homogeneous antiferromagnetic state.
Equivalently, the layer-resolved topological charge changes sequentially.
The initial SAF pair has one skyrmion charge in each layer, with opposite signs for the antiferromagnetically
coupled textures.
At the first saddle the lower-layer charge changes to zero, while the upper-layer charge remains nonzero.
The intermediate therefore carries topological charge only in the upper layer.
At the second saddle this remaining charge is also lost, and the final pinned collinear state has zero
topological charge in both layers.

\begin{figure}
\centering
\includegraphics[width=\linewidth]{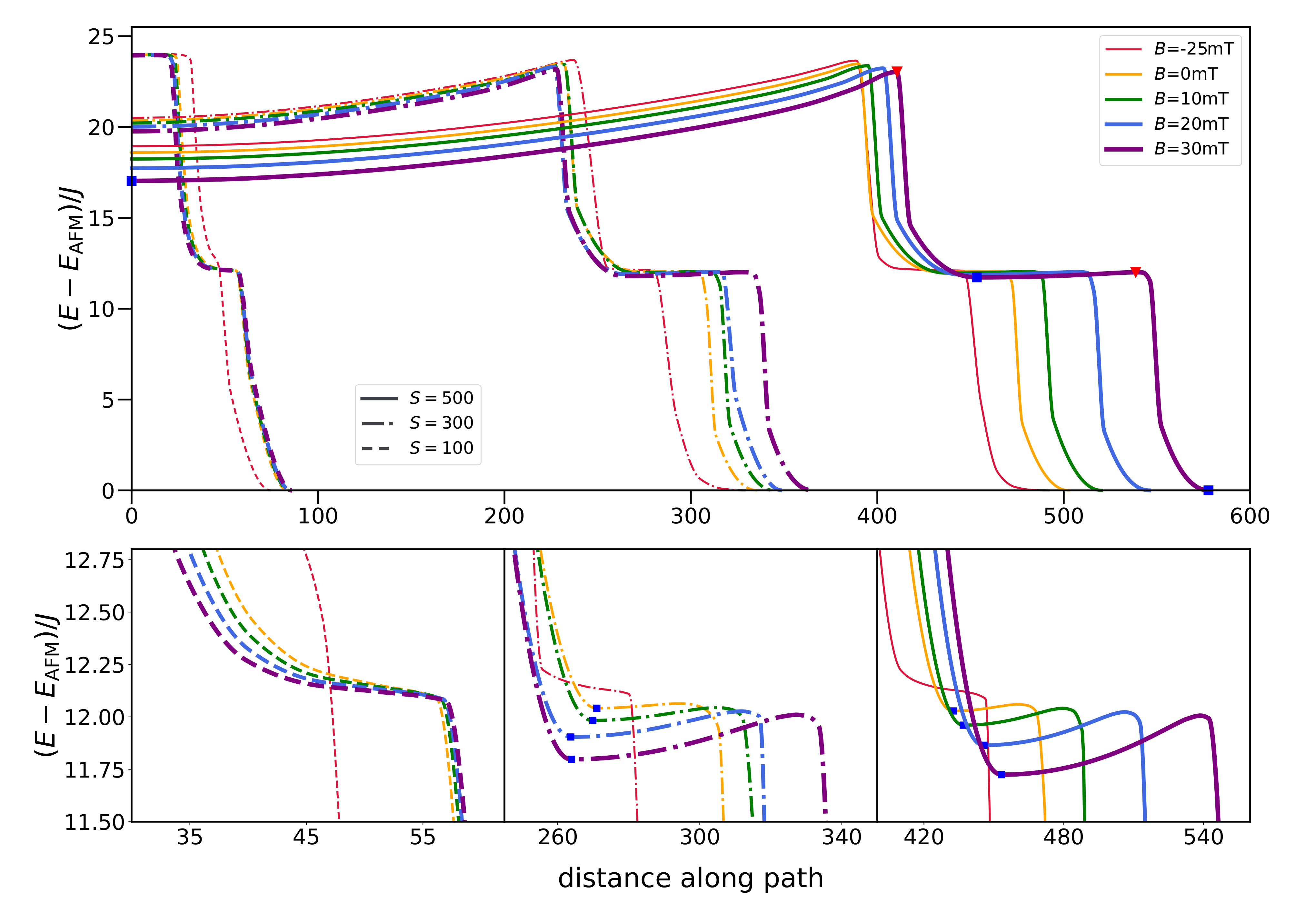}
\caption{Minimum energy paths for skyrmion-pair collapse at different applied fields $B$ and pinned-island sizes
$S$.
Energies are measured relative to the pinned homogeneous antiferromagnetic reference state.
Representative local minima and saddle points are marked by squares and triangles, respectively.
The lower insets enlarge the single-layer-skyrmion segment; from left to right, they correspond to $S=100$,
$300$, and $500$.
In all panels, color and line width code $B$, while the dash pattern codes $S$ (dashed $100$, dash-dotted
$300$, solid $500$).
The intermediate minimum is the upper-layer skyrmion.
}
\label{fig:meps}
\end{figure}

The same structure is seen in the full set of MEPs in Fig. \ref{fig:meps}.
The path starts from the skyrmion-pair minimum at positive energy relative to the pinned collinear reference
state.
It then climbs to the main saddle point, drops to the energy range of the upper-layer skyrmion, crosses a
lower saddle point, and finally reaches the homogeneous state.
The position along the horizontal axis is the accumulated Euclidean distance in the full spin configuration
space; it is therefore larger for larger pinned islands and should not be interpreted as a real-space coordinate.

The reverse process follows the same MEP in the opposite direction.
Starting from the pinned collinear reference state, the upper-layer skyrmion is nucleated first, and only then
is the lower-layer partner nucleated.
When this single-layer state is a local minimum, it is a genuine intermediate on the path.

\subsection{Interior collapse barrier and role of confinement}

\begin{figure}
\centering
\includegraphics[width=\linewidth]{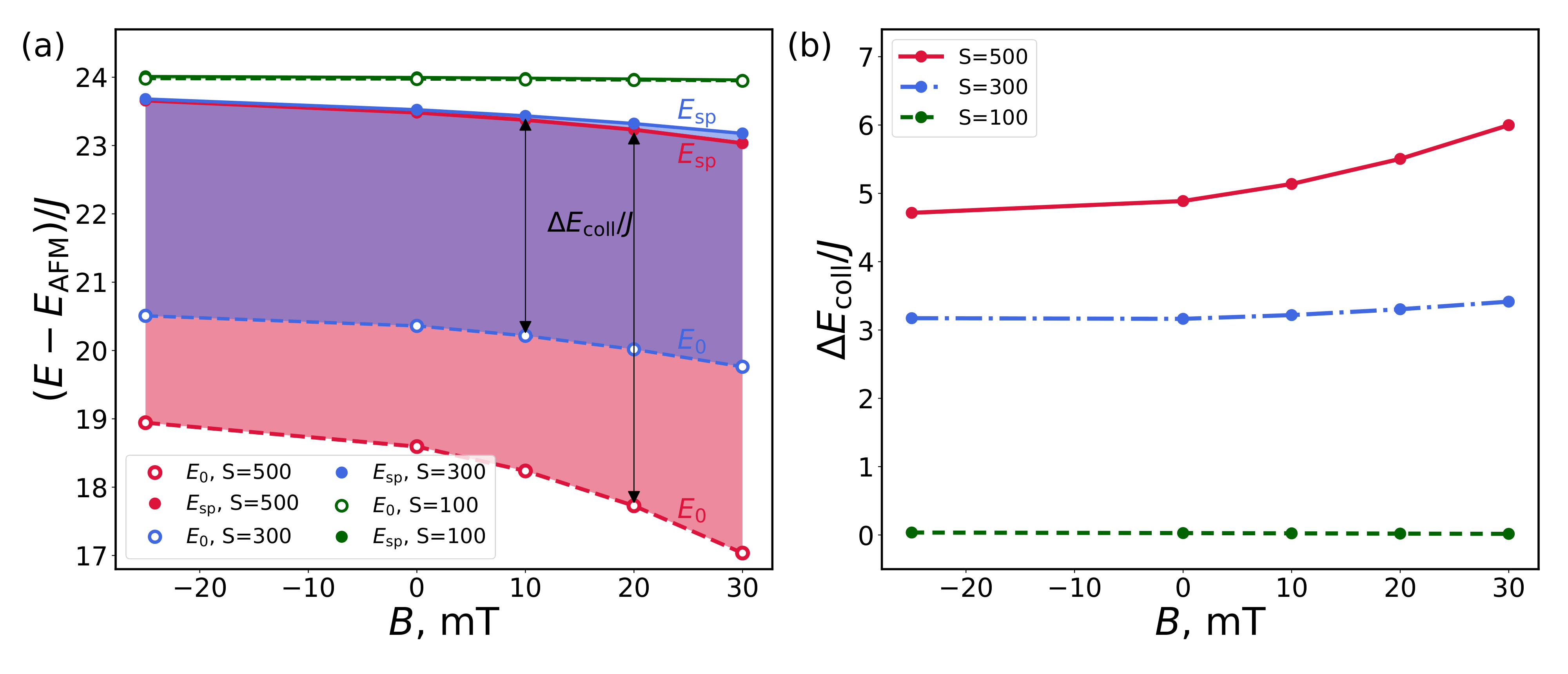}
\caption{(a) Energy of the skyrmion-pair minimum $E_0$ and the main saddle point $\Esp$ relative to the
pinned collinear antiferromagnetic reference state as a function of the applied field $B$.
(b) Pinned-boundary interior collapse barrier $\Dcoll=\Esp-E_0$ for different pinned-island sizes $S$.
In panel (b), line styles encode $S$ as in Fig. \ref{fig:meps}.
In this pinned-boundary calculation, the size dependence of the barrier is dominated by the strong change of
the skyrmion-pair minimum $E_0$, while the main saddle energy varies much more weakly.}
\label{fig:barrier_c}
\end{figure}

Figure \ref{fig:barrier_c} shows the energy of the skyrmion-pair minimum $E_0$ and the energy of the main
saddle point $\Esp$, together with the resulting collapse barriers for this MEP channel.
The skyrmion pair is a local minimum in all cases shown, although its energy lies above the pinned collinear
reference state.
The two energy branches have different size dependences.
The main saddle energy varies comparatively weakly with $S$ and remains in a narrow interval near
$23$--$24J$.
By contrast, the absolute energy of the skyrmion pair depends strongly on pinned-island size and should be read
as an upper-bound estimate that includes the boundary penalty imposed by the fixed collinear edge.
For $S=100$, $E_0$ is close to $24J$ and is nearly degenerate with the saddle point, leading to an almost
vanishing collapse barrier.
For $S=300$, $E_0$ lies near $20J$ and the barrier is about $3.1$--$3.4J$.
For $S=500$, $E_0$ decreases from about $19J$ at $B=-25$ mT to about $17J$ at $B=30$ mT, while the saddle
energy remains near $23J$.
As a result, the barrier of this channel increases from about $4.7J$ to about $6.0J$.

The main observation from Fig. \ref{fig:barrier_c} is that the pinned-boundary energy barrier for collapse
$\Dcoll=\Esp-E_0$ is controlled mainly by the skyrmion-pair energy rather than by a variation of saddle energy.
Increasing $B$ also raises the barrier by lowering the skyrmion-pair energy, but over the
explored interval its effect is moderate compared with the change of lateral size from $S=100$ to $S=500$.
The smallest pinned island is an important limiting case: the relaxed pair exists as a local configuration, but the
saddle point is so close in energy that the pair has essentially no barrier protection against thermal
collapse.
For an open finite island, the corresponding barrier would have to be recalculated with boundary relaxation and
possible edge-assisted paths.

The field-induced increase of the barrier is almost absent for $S=100$ and becomes noticeable only for
$S\gtrsim 300$.
Thus, over the explored range, an applied field acts as a secondary tuning parameter compared with lateral
confinement.

In the reduced model, this result gives a qualitative estimate for the calculated interior collapse channel.
In HTST, the lifetime of a metastable state has the Arrhenius form
\begin{equation}
\tau \simeq \tau_0\exp\left(\frac{\Dcoll}{k_{\mathrm B}T}\right),
\end{equation}
where $\tau_0$ contains the dynamical and entropic prefactors.
With the parameter mapping used here, $J=183.5$ meV.
Representative reduced barriers of $3J$, $5J$, and $6J$ therefore correspond to approximately $0.55$, $0.92$,
and $1.10$ eV, or about $21$, $35$, and $43$ in units of $k_{\mathrm B}T$ at $T=300$ K
($k_{\mathrm B}T\simeq25.9$ meV).
These numbers are useful for interpreting the exponential factor of the pinned-boundary reduced Hamiltonian,
but an absolute lifetime would also require the Hessian and dynamical prefactors, the omitted anisotropy and
magnetostatic contributions, and a check for competing edge-assisted paths.
For a given material and operating temperature, this gives the formal Arrhenius scale condition:
\begin{equation}
\Dcoll(S,B) \gtrsim k_{\mathrm B}T\ln\left(\frac{\tau_{\mathrm{ret}}}{\tau_0}\right),
\end{equation}
where $\tau_{\mathrm{ret}}$ is the target retention time.
The figures show the dimensionless value $\Dcoll/J$; multiplying it by the material exchange scale $J$ gives
the energy scale of this pinned-boundary model barrier.
Thus the present barrier map gives the Arrhenius exponent of the calculated interior collapse channel.
The inequality is used here as a scale comparison for the exponential part of the rate; converting it into a
device retention criterion requires the additional ingredients listed above.

\subsection{Reverse barrier and assisted writing}

\begin{figure}
\centering
\includegraphics[width=\linewidth]{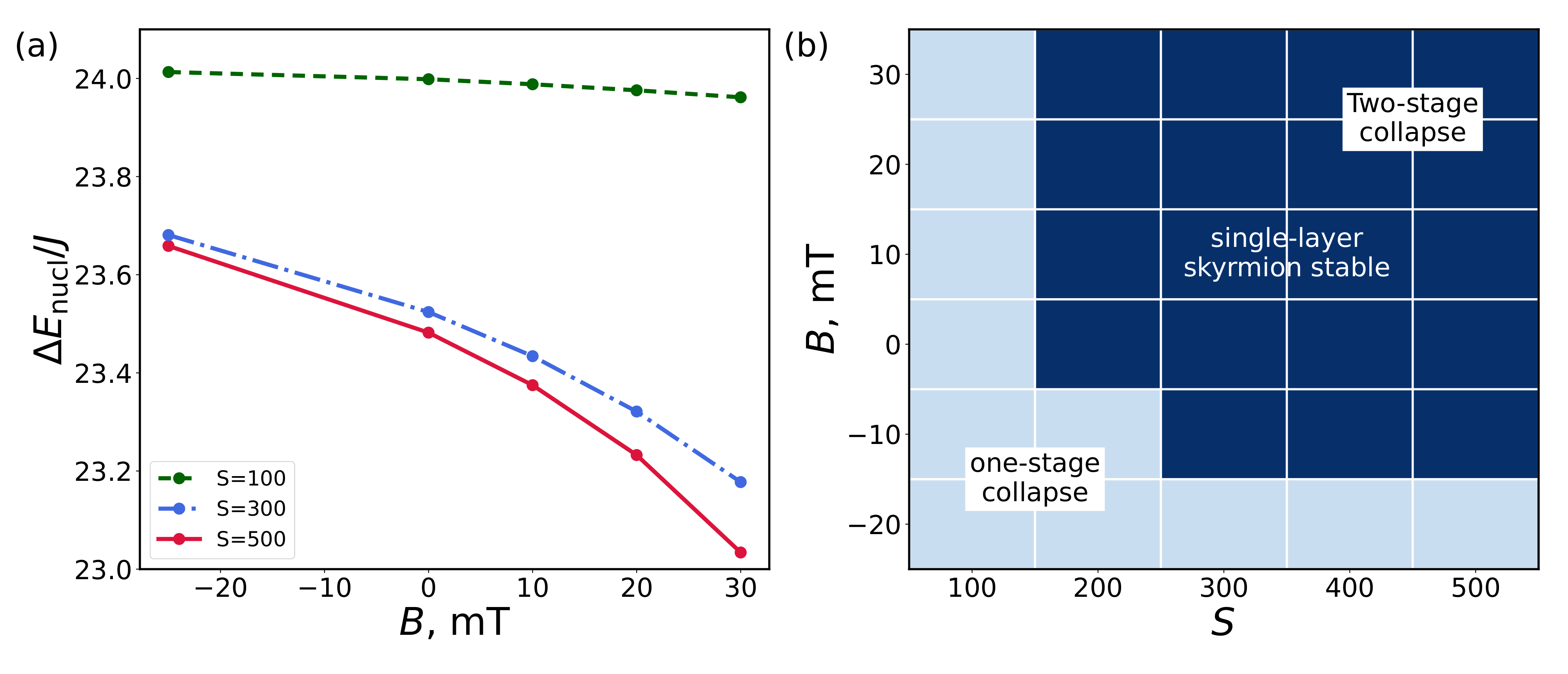}
\caption{(a) Reverse barrier $\Dnucl$ from the pinned collinear antiferromagnetic reference state to the bound
skyrmion pair along the same minimum energy path; line styles encode $S$ as in Fig. \ref{fig:meps}.
(b) Stability diagram for the upper-layer ($l=2$) single-skyrmion intermediate as a function of applied field $B$
and pinned-island size $S$.
The bound skyrmion pair satisfies the relaxation-based local-stability criterion throughout the displayed
region.
Each colored cell represents one calculated grid point; no interpolation between grid points is used.
The darker cells mark parameters where the upper-layer single-skyrmion state also satisfies this criterion:
the state remains after independent nonlinear conjugate-gradient relaxation, the optimized MEP contains a
strict intermediate minimum, and the minimum is deeper than $5\times10^{-4}J$ relative to the neighboring
images.
In these cells the collapse/nucleation path is genuinely two-stage, but the seed-decay barrier
$E_{\mathrm{sp},2}-E_{\mathrm{1sk}}$ remains small, $0.022$--$0.282J$, compared with barrier for completing the pair
$\Delta E_{\mathrm{comp}}\approx 11J$.}
\label{fig:barrier_n}
\end{figure}

The reverse transition has a very different energy scale.
Figure \ref{fig:barrier_n} shows the reverse barrier measured relative to the pinned collinear
antiferromagnetic reference state.
It is approximately $23$--$24J$ over the whole parameter range, much larger than the collapse barrier of an
existing skyrmion pair.
The barrier decreases slightly with increasing positive applied field for the larger pinned islands, but the change is
small compared with the absolute value of the barrier.

The numerical values underlying Figs. \ref{fig:meps}--\ref{fig:barrier_n} are listed in
Table \ref{tab:mep_summary}. Here $E_{\mathrm{sp},1}$ is the main, higher saddle point, $E_{\mathrm{sp},2}$ is the lower saddle, and
$\Delta E_{\mathrm{comp}}=E_{\mathrm{sp},1}-E_{\mathrm{1sk}}$ is the barrier for completing the pair from
intermediate minimum.
The competing decay barrier of the intermediate is $E_{\mathrm{sp},2}-E_{\mathrm{1sk}}$.

\begin{table}[tbp]
\centering
\caption{Numerical summary of the calculated MEPs. $L=Sa$ is the lateral size in nanometers. Energy columns are
in units of $J$ relative to the pinned collinear reference state. A dash means that no accepted upper-layer single-skyrmion minimum is present on the final
discrete path.}
\label{tab:mep_summary}
{\footnotesize
\setlength{\tabcolsep}{0pt}
\renewcommand{\arraystretch}{1.08}
\begin{tabular*}{\textwidth}{@{\extracolsep{\fill}}rrrcrrrrrrr@{}}
\hline
$S$ & $L$ (nm) & $B$ (mT) & $N_{\mathrm{img}}$ & $E_0/J$ & $E_{\mathrm{sp},1}/J$ &
$\Dcoll/J$ & $E_{\mathrm{1sk}}/J$ & $E_{\mathrm{sp},2}/J$ & $\Delta E_{\mathrm{comp}}/J$ &
$\Dnucl/J$ \\
\hline
100 & 25.1 & -25 & 20 & 23.977 & 24.013 & 0.036 & -- & -- & -- & 24.013 \\
100 & 25.1 & 0 & 20 & 23.971 & 23.998 & 0.028 & -- & -- & -- & 23.998 \\
100 & 25.1 & 10 & 20 & 23.964 & 23.988 & 0.024 & -- & -- & -- & 23.988 \\
100 & 25.1 & 20 & 20 & 23.956 & 23.976 & 0.020 & -- & -- & -- & 23.976 \\
100 & 25.1 & 30 & 20 & 23.945 & 23.962 & 0.017 & -- & -- & -- & 23.962 \\
300 & 75.2 & -25 & 23 & 20.508 & 23.681 & 3.173 & -- & -- & -- & 23.681 \\
300 & 75.2 & 0 & 32 & 20.360 & 23.524 & 3.164 & 12.042 & 12.064 & 11.482 & 23.524 \\
300 & 75.2 & 10 & 32 & 20.215 & 23.434 & 3.219 & 11.983 & 12.045 & 11.451 & 23.434 \\
300 & 75.2 & 20 & 32 & 20.017 & 23.321 & 3.304 & 11.904 & 12.027 & 11.417 & 23.321 \\
300 & 75.2 & 30 & 37 & 19.762 & 23.177 & 3.415 & 11.798 & 12.010 & 11.380 & 23.177 \\
500 & 125.3 & -25 & 37 & 18.944 & 23.659 & 4.715 & -- & -- & -- & 23.659 \\
500 & 125.3 & 0 & 32 & 18.594 & 23.482 & 4.888 & 12.028 & 12.060 & 11.454 & 23.482 \\
500 & 125.3 & 10 & 27 & 18.237 & 23.375 & 5.138 & 11.961 & 12.041 & 11.414 & 23.375 \\
500 & 125.3 & 20 & 34 & 17.728 & 23.232 & 5.505 & 11.865 & 12.023 & 11.368 & 23.232 \\
500 & 125.3 & 30 & 34 & 17.035 & 23.034 & 5.999 & 11.724 & 12.006 & 11.310 & 23.034 \\
\hline
\end{tabular*}
}
\end{table}

The single-layer intermediate is a shallow state.
Where it is present, its decay barrier $E_{\mathrm{sp},2}-E_{\mathrm{1sk}}$ is only $0.022$--$0.282J$.
With $J=183.5$ meV, this corresponds to about $0.16$--$2.0\,k_{\mathrm B}T$ at room temperature.
By contrast, the barrier for completing the pair from the same intermediate,
$\Delta E_{\mathrm{comp}}$, is about $11J$.
Thus, in an undriven thermal process, the single-layer state would be expected to decay over the lower saddle
much more readily than to complete the pair.
For writing, it marks a configuration that an external drive could create, maintain, or rapidly convert before
it relaxes back to the pinned collinear state.

Within the calculated pinned-boundary MEPs, the barrier for nucleation is much larger than the interior collapse barrier in
Fig. \ref{fig:barrier_c}, implying a strong activation-barrier asymmetry between formation from the pinned
reference state and this particular decay channel of an already written skyrmion bit. A large $\Dnucl$ raises the Arrhenius exponent for spontaneous thermal formation from the pinned collinear
state, which is desirable if this state represents the unwritten bit.
The same large barrier also means that direct thermal nucleation of a complete SAF pair would be inefficient in
this reduced model.
Writing through this path would require an external drive, such as current, voltage, laser heating, or a
geometric injector, to bypass or reshape the equilibrium energy barrier.

The two-stage path clarifies what such a drive should accomplish.
If the system starts from the pinned collinear reference state and no external drive is maintained, the overall
rare-event exponent along this path is still set by the highest saddle point relative to $\EAFM$.
The intermediate minimum alone does not reduce the equilibrium reverse-path exponent.
However, if the upper-layer skyrmion is deliberately created, or injected into the SAF region from a
single-layer part of the device, the relevant second step starts from the upper-layer skyrmion minimum rather
than from the pinned collinear reference state.
The barrier for completing the pair is then the energy difference between the high saddle point and the
single-layer minimum.
In the paths shown here this is roughly half of the full static reverse barrier.
This lower second-step barrier motivates layer-selective or layer-sequential writing.

\subsection{Stability of the upper-layer intermediate}

The intermediate upper-layer skyrmion does not always satisfy the relaxation-based stability criterion.
Figures \ref{fig:meps} and \ref{fig:barrier_n}(b) show enlarged parts of the MEPs and the corresponding
relaxation-based stability map near the single-layer segment.
For $S=100$ the path contains no accepted local minimum in this region.
After the lower-layer skyrmion collapses, the remaining upper-layer skyrmion is unstable, and the path
continues directly toward the homogeneous state.
In this sense, the collapse is effectively single-stage.

For $S=300$, the upper-layer single-skyrmion state passes the relaxation criterion for most of the explored fields,
although the minimum is shallow and absent or marginal at the most negative field.
The depth of the minimum shifts with $B$, indicating that the applied field controls the stability
of the upper-layer seed, even though it has a weaker effect on the total collapse barrier than the lateral size.

For $S=500$, the local minimum is well developed for nonnegative fields, while the $B=-25$ mT path has no
accepted intermediate minimum in the final discrete MEP.
Where the minimum is present, the two-stage character of the path is well resolved: the pair first loses one of
its two skyrmions, then the upper-layer skyrmion survives as a relaxed local configuration, and only after
crossing the lower saddle point does the system become homogeneous.

The resulting relaxation-based stability map is summarized in Fig. \ref{fig:barrier_n}(b).
The bound skyrmion pair satisfies this criterion in the whole parameter range considered.
The upper-layer single-skyrmion state satisfies it only in the darker region.
Operationally, a darker cell means that the independently relaxed upper-layer state remains stationary with
maximum tangent-space gradient below $10^{-9}J$, appears as a strict discrete MEP minimum, and has a depth of at
least $5\times10^{-4}J$ relative to the neighboring path images.
The state is absent in the smallest pinned islands, appears at nonnegative or weakly negative applied fields for
intermediate sizes, and passes the relaxation criterion over a wider field range as the pinned-island size increases.
However, Fig. \ref{fig:barrier_n}(b) is an existence map of the local minimum rather than a passive-retention
map for the seed.
For every dark cell in the present parameter set, the barrier for losing the single-layer skyrmion,
$E_{\mathrm{sp},2}-E_{\mathrm{1sk}}$, is only $0.022$--$0.282J$.
This corresponds to $0.16$--$2.0\,k_{\mathrm B}T$ at room temperature and is much smaller than
$\Delta E_{\mathrm{comp}}=E_{\mathrm{sp},1}-E_{\mathrm{1sk}}\simeq 11.3$--$11.5J$, the barrier for completing
the pair from the same intermediate.

For writing, Fig. \ref{fig:barrier_n}(b) complements the barrier plot.
A layer-by-layer writing protocol requires the first skyrmion to be maintained long enough to seed the second layer.
For the field orientation used here, this corresponds to first creating the upper-layer skyrmion and then the
lower-layer partner.
The dark region of the stability diagram identifies where such a state exists as a relaxed local minimum; the
small decay barrier makes it an actively maintained intermediate rather than a thermally robust seed.
In the light region, an upper-layer seed is not metastable; a writing pulse would need to create the second
skyrmion before the first one collapses, or maintain the seed dynamically throughout the process.

\section{Model implications for SAF skyrmion memory}

The calculated energy landscape of the reduced Hamiltonian distinguishes two quantities: the pinned-boundary
interior collapse barrier and the reverse barrier for nucleation from the pinned collinear reference state.
For fixed reduced-model parameters and this boundary condition, the pinned-island size is
the dominant control parameter for the calculated interior-collapse channel.
The size dependence enters mainly through the energy of the skyrmion-pair minimum: the interior saddle energy changes
comparatively weakly, but the pair minimum is strongly affected by the fixed boundary.
The smallest pinned island therefore has an almost zero barrier despite containing a relaxed skyrmion-pair
configuration, whereas the larger pinned islands develop channel barriers of several exchange energies.
Increasing the applied field raises the barrier in the explored range, but it does not compensate for strong
lateral confinement.
Thus a stationary relaxed pair in a small pinned island is insufficient to establish useful barrier-limited
stability.

Changing the magnetic layer thickness is another parameter that could affect the barrier scale, but its effect
is ambiguous.
In a simple scaling picture, if the dimensionless texture and all coupling ratios remain fixed, $J\propto d$
and the activation barriers scale approximately linearly with the magnetic thickness.
In that limit a thicker layer increases the barrier exponent without changing the shape of the MEP.
In a real SAF, however, RKKY coupling is an interfacial effect controlled mainly by the spacer and interfaces.
Increasing the magnetic thickness can therefore reduce $J_{12}/J$ and the fixed-bias ratio in the effective
lattice model, even while increasing the absolute exchange scale.
Thickness changes should consequently be viewed as a trade-off between increasing the energy scale and
preserving the effective interlayer coupling needed to bind the two skyrmions.
Because thickness, anisotropy, and dipolar interactions are not taken into account explicitly in (\ref{eq:Hamiltonian}), the calculations provide a
map of the retained effective parameters rather than an optimization of a real stack.
In particular, no control MEP with an explicit perpendicular anisotropy $K$ or with magnetostatic fields was
performed.
These terms can change the skyrmion radius, the local-stability region of the single-layer state, the absolute
pair and saddle energies, and the numerical lifetime estimate in Pt/Co/Ru-based SAF stacks.
The structural trends are the most likely to survive these added terms: once an antiferromagnetically bound
pair exists, interlayer exchange and layer asymmetry can make the transition layer-resolved; the reverse static
MEP can pass through a one-layer configuration; and this configuration can be useful for actively driven
layer-by-layer writing.
By contrast, the quantitative values of $E_0$, $\Esp$, the decay barrier of the one-layer state,
$\Delta E_{\mathrm{comp}}$, and the size threshold for useful stability should be recalculated in a model that
includes perpendicular anisotropy and demagnetizing fields.
Because SAF compensation reduces the long-range stray field but does not remove residual layer-imbalance and
edge magnetostatics, such a calculation is also needed to decide whether a competing edge-assisted path becomes
the lowest decay route in a real device geometry.

For writing, the relevant process is actively driven.
Assisted writing of SAF skyrmions has already been studied by current pulses, laser excitation, voltage
control, thermal effects, and hybrid FM/SAF geometries \cite{Juge_22,Qiu22,Chen_22,Deng23}.
In the pinned-boundary reduced model, the MEP connects the bound pair and the pinned collinear reference state
through an upper-layer single-skyrmion intermediate.
Therefore, for the present field orientation, a writing scheme that first creates the upper-layer skyrmion and
then uses the interlayer exchange to generate its lower-layer partner is consistent with this reduced-model
energy landscape, provided the drive maintains or rapidly converts the single-layer state.

A similar idea appears in hybrid FM/SAF proposals \cite{Deng23}.
For example, one can consider an asymmetric island in which one ferromagnetic layer extends beyond the overlap
region.
A skyrmion is first nucleated in the single-layer or weakly compensated part, where layer-selective
manipulation is easier, and then driven into the bilayer SAF region.
Once inside the overlap region, the seed skyrmion produces a nonuniform exchange field that acts on the second
layer and can assist the formation of the antiferromagnetically coupled partner skyrmion.
The MEP calculations add that, in the explored parameter range, the upper-layer skyrmion satisfies the
relaxation-based local-minimum criterion only above a size-dependent threshold, and even there its own decay
barrier is small.
They also show that completing the pair has a much smaller bottleneck when an external drive prepares and
supports the upper-layer seed than when the system nucleates the full pair spontaneously from the pinned
collinear reference state.

\section{Conclusions}

Minimum energy paths for pinned SAF islands show that the pinned-boundary activation energy for interior
collapse along the calculated MEP channel is strongly affected by lateral confinement.
This size trend is dominated by the skyrmion-pair minimum, whose energy carries the boundary penalty, rather
than by a comparable shift of the localized saddle energy.
In the parameter range studied here, the resulting barrier is nearly zero for $S=100$, reaches about $3J$ for
$S=300$, and increases to about $5$--$6J$ for $S=500$.
Using $J=183.5$ meV, these representative barriers correspond to roughly $0.55$ eV and
$0.92$--$1.10$ eV, or about $21$ and $35$--$43$ in units of $k_{\mathrm B}T$ at room temperature.
The applied field also increases the barrier by lowering the skyrmion-pair energy, but its influence is weaker
than that of the lateral size.
Thus the smallest pinned island contains a relaxed stationary pair with negligible protection along the calculated
interior-collapse channel, whereas the larger pinned islands develop channel barriers of several exchange energies.

This boundary condition matters, but the size trend remains clear.
The pinned edge represents a fixed homogeneous antiferromagnetic background and suppresses edge twisting and
skyrmion escape.
It also raises the absolute energy of finite skyrmion-pair textures, so pair energies should be regarded as
upper-bound estimates relative to boundary-relaxed calculations.
The saddle energies, by contrast, characterize the localized interior collapse mechanism targeted by the MEP.
The reported barriers therefore give the Arrhenius exponent for this reduced pinned-boundary channel; full
lifetimes and device-level retention estimates require the omitted anisotropy, magnetostatic terms, HTST
prefactors, and possible competing edge-assisted paths.
Other decay routes, including edge-assisted ones, require separate calculations.

For the field orientation used here, the larger pinned islands show a layer-sequential transition.
The lower-layer skyrmion collapses first, leaving an upper-layer skyrmion minimum when that state satisfies the
relaxation criterion, and only then does the remaining skyrmion collapse.
The reverse nucleation path is therefore also layer-sequential.
The stability map shows where this upper-layer intermediate exists as a relaxed local minimum.
However, its own decay barrier is small, about $0.02$--$0.28J$ in the accepted cases, so it is better viewed
as an actively supported intermediate for layer-by-layer writing than as a thermally robust seed.

The reverse barrier for nucleation from the pinned collinear antiferromagnetic reference state is much larger than the
collapse barrier of an already written pair, about $23$--$24J$ over the explored parameter range.
This asymmetry makes spontaneous formation along the calculated path unlikely and points to assisted writing
rather than direct thermal nucleation.
In particular, if the upper-layer skyrmion is created or injected first, interlayer exchange can complete the
bound SAF pair through a much smaller second-step barrier, provided the nonequilibrium drive acts before the
single-layer seed decays.
Overall, the MEPs separate two roles of the energy landscape: the interior-collapse barrier
quantifies the size trend of the calculated decay channel, while the upper-layer intermediate provides a candidate
state for actively driven layer-by-layer writing with this field orientation.


{ \setlength{\parindent}{0mm} \setlength{\parskip}{3mm}
\textit{Information about the author:}

\textit{Maria N. Potkina} -- Infochemistry Scientific Center, ITMO University, 197101 St.
Petersburg, Russia; ORCID 0000-0002-1380-2454; potkina.maria@yandex.ru

\textit{Conflict of interest}: the author declares no conflict of interest.
}

\end{document}